\def\year{2019}\relax
\documentclass[letterpaper]{article} 
\usepackage{aaai19}  
\usepackage{times}  
\usepackage{helvet}  
\usepackage{courier}  
\usepackage{url}  
\usepackage{graphicx}  
\usepackage{amsmath}
\usepackage{amsfonts}
\usepackage{bm}
\usepackage{color}
\usepackage{setspace}
\usepackage{subcaption}
\usepackage{booktabs}
\usepackage{mathtools}
\usepackage{multirow}
\usepackage{float}
\usepackage{caption}
\usepackage{adjustbox}
\usepackage{cleveref}
\usepackage{makecell}
\usepackage{tabularx}

\crefname{equation}{Eq.}{Eq.}
\crefname{section}{Section}{Sections}
\crefname{subsection}{Section}{Sections}
\crefname{subsubsection}{Section}{Sections}
\crefname{figure}{Figure}{Figures}
\crefname{table}{Table}{Tables}
\crefname{subfigure}{Figure}{Figures}
\crefname{algocf}{Algorithm}{Algorithms}

\begin{document}
\title {Beyond ``How may I help you?'': \\ 
Assisting Customer Service Agents with Proactive Responses}

\author{
Mengting Wan\\
University of California, San Diego\\
m5wan@ucsd.edu\\
\And 
Xin (Cindy) Chen\\
Airbnb, Inc.\\
cindy.chen@airbnb.com\\
}
\maketitle
\begin{abstract}
We study the problem of providing recommended responses to customer service agents in live-chat dialogue systems.
Smart-reply systems have been widely applied in real-world applications (e.g.~Gmail, LinkedIn Messaging), where most of them can successfully recommend \textit{reactive} responses.
However, we observe a major limitation of current methods is that they generally have difficulties in suggesting \textit{proactive} investigation act (e.g.~``Do you perhaps have another account with us?'') due to the lack of long-term context information, which indeed act as critical steps for customer service agents to collect information and resolve customers' issues. Thus in this work, we propose an end-to-end method with special focus on suggesting proactive investigative questions to customer agents in Airbnb's customer service live-chat system. Effectiveness of our proposed method can be validated through qualitative and quantitative results.
\end{abstract}

\section{Introduction}

\begin{figure}[t]
    \centering
	\includegraphics[width=0.8\linewidth]{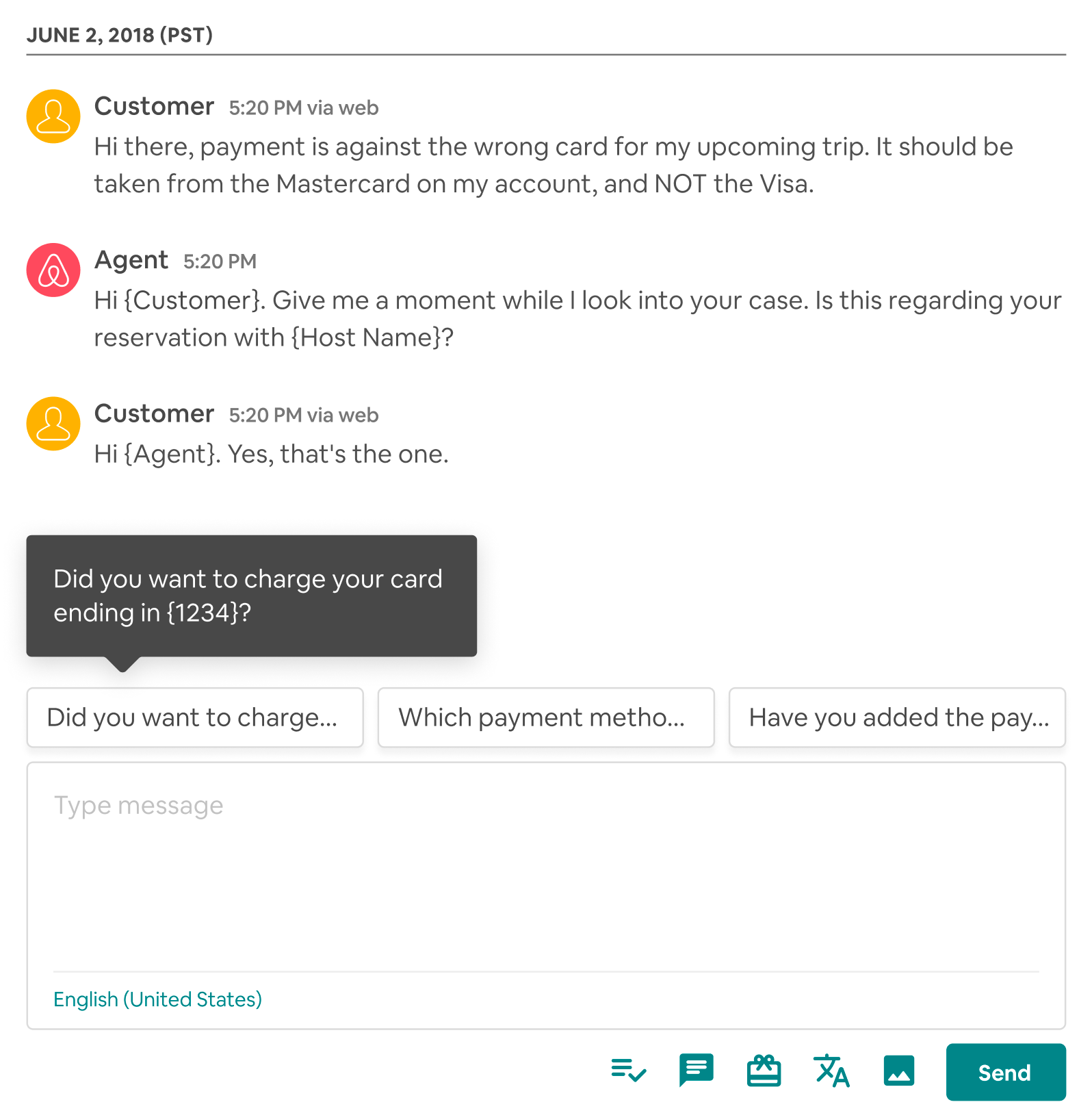}
	\caption{\small An example of dialogue assistant for Airbnb's live-chat customer service agents. The recommendations carry the context from previous rounds: customer's intent to change payment method. (Placeholders such as customer, agent, host name, and 1234 are auto-filled in real applications.)}
\end{figure}

Smart-reply systems have been adopted in many real-world applications and have assisted numerous textual responses in people's day-to-day email and messaging activities (e.g. Gmail, LinkedIn Messaging). These systems are particularly effective in reducing users' cognitive load of processing text content and the inconvenience of typing responses. 

At Airbnb, we serve the community of hosts and guests around the world. It is an essential part of Airbnb's business to provide high quality customer service at scale. Live-chat is one of the primary customer support channels, where customers work with our customer service agents to resolve their issues through live-chat conversations. We observed that it was time-consuming for our live-chat agents to repeatedly type or copy-paste commonly used messages in their conversations. It could become an overhead preventing the agents from deep engagement with the customers. These observations initially motivated us to introduce `smart replies' into our customer support live-chat conversations, where we wish to suggest short responses for agents to use with 1 click. 

Most existing methods of `smart replies' seek to build a sequence-to-sequence or a classification model on top of the preceding message and generate the corresponding response \cite{kannan2016smart,henderson2017efficient,linkedinblog}. We first deployed a variant of such systems in Airbnb's live-chat customer service and evaluated its performance. This first version of smart replies has resulted in significant positive impact in agent efficiency metrics and received positive feedback from our agents. However, we also observed a unique challenge in our application scenario. Unlike existing smart-reply systems where messages involved are mostly asynchronous and reactive, in a live-chat conversation, there are many rounds of back and forth between the customer and the agent (see  \cref{table:stats}). Apart from reacting to customers' preceding inquires, agents may ask questions to investigate customers' problems and provide information to resolve customer issues. Thus in order to improve the efficiency of the problem-solving process as well as providing consistent investigation guidance to our agents, we extend our work to focus on suggesting \textit{proactive investigative questions} to Airbnb's live-chat customer service agents.

\begin{table}
	\centering
	\small
	\begin{tabular}{rcc}
		\toprule
		\textbf{} & \textbf{Median} & \textbf{Mean} \\
		\midrule
		\textbf{Number of Messages } & 17.4 & 14 \\
		\midrule
		\textbf{Number of Turns} & 7.9 & 7.0 \\
		\midrule
		\textbf{Number of Rounds} & 4.7 & 4 \\
		\bottomrule
	\end{tabular}
	\caption{\small Median and mean number of messages, turns and rounds in live-chat conversations. One round contains two consecutive turns from the agent and the customer, exceptions at start and end} \label{table:stats}
\end{table}

\section{Related Work}

Modeling conversation dialogues and generating responses have been extensively studied recently. Most relevant to our work are smart-reply systems, where the preceding message is taken as the input and the corresponding response is generated accordingly \cite{vinyals2015neural,kannan2016smart,henderson2017efficient,linkedinblog}. Specifically, given the sequence of tokens in the most recent message, the conditional probability of response tokens can be regarded as the output sequence, and text generation \cite{kannan2016smart,vinyals2015neural} or classification \cite{henderson2017efficient,linkedinblog} models can be used to approximate such a probability distribution. At the inference time, response candidates with high probabilities will be suggested.

We wish to extend existing smart-reply systems to customer service live-chat dialogues. Notice that in the above described systems, only the \textit{short-term} memory (i.e.,~the preceding message) is involved as the input sequence. We observe that such a method is particularly suitable for suggesting \textit{reactive} responses and answer customer questions.  

\begin{itemize}
    \item \textbf{Reactive Responses.} The intention of this type of messages is promptly responding to the current situation rather than leading the conversation flow, which largely relies on the short-term memory only. For example, it is relatively straightforward to predict ``You're welcome!'' or ``It's my pleasure.'' by looking at the most recent message ``Thank you!''.
    \item \textbf{Question and Answer.} Another type of tasks these short-term methods are good at is to answer customers' simple questions (e.g.~questions do not require further investigations and can be answered in one single round). For example, answering ``Your payout will be released in 24 hours.'' to the question ``When will my payout be released?''. Again, this use case is reactive in nature and mostly relies on the short-term memory. 
\end{itemize}
However, we find existing methods have difficulties in suggesting common responses that include investigation intents (e.g.~``Can you provide the reservation code?'', ``Do you perhaps have a duplicated account with us?''). Different from previous reactive responses, these messages are \textit{proactive} (i.e., it requires agents to lead conversations and investigation flows), \textit{dynamic} (i.e., solving a problem could require a sequence of investigative questions), and \textit{issue-dependent} (i.e., different types of problem issues may lead to different resolving processes). Thus in order to suggest these investigative questions, our system needs to be capable of capturing the short-term context from the preceding massage, as well as understanding the long-term problem context throughout the conversation.

On the other hand, chatbot systems have raised much attention in the customer service domain \cite{xu2017new,cui2017superagent}. Our work differs from these systems in terms of the following two aspects: 1) unlike most chatbot systems which utilize language generation models to predict the full responses, we develop a pipeline to efficiently generate and carefully curate the response candidates, and regard the ultimate response suggestion task as a candidate ranking problem; 2) different from the chatbot system \cite{cui2017superagent} where the problem-solving task is regarded as a single-round question-answering problem, our smart-reply system is designed to solve more sophisticated user issues where long-term problem contexts are required to be considered.

Many other studies exists on modeling dialogues, such as \cite{serban2016building,sordoni2015neural,shang2015neural,li2016deep,li2016diversity}. Although most of these models have not been deployed in production settings, some of them did inspire us in terms of the model architecture design. Specifically, our model is largely inspired by the \textbf{HRED} architecture \cite{serban2016building} where hierarchical recurrent encoder-decoders are applied to capture both short-term and long-term conversational contexts.

In this work, we focus on suggesting \textit{investigative questions} to agents. In the subsequent section, we will describe 1) how we define and uncover investigative question candidates from Airbnb's agent-customer conversation data; and 2) how we build a machine learning system to suggest these response candidates to agents by leveraging both short-term conversation contexts and long-term problem-solving contexts.

\section{Method}
We describe a typical live-chat workflow as follows. A user first describes the problems and this initializes a customer \textit{ticket}, which will be routed to one of our live-chat agents. The live-chat agent picks up the ticket afterwards and determines the associated \textit{ticket issue} category (e.g.~ ``cancel reservation'', ``change payment method'').
This ticket issue category can be selected by the customer when filing the ticket, re-assigned by an agent when picking up the ticket, or remaining blank (i.e.,~missing) throughout the conversation. Then the agent-customer live-chat \textit{conversation} starts with the original customer problem description message, and keeps running for several conversation \textit{rounds} in real-time until the ticket is solved or escalated to another channel. 

Then we introduce our system from the following two perspectives: \textit{investigative query candidate generation}, and \textit{investigative question recommendation}. 

\subsection{Investigative Question Candidate Generation}

\begin{figure*}
	\includegraphics[width=\linewidth]{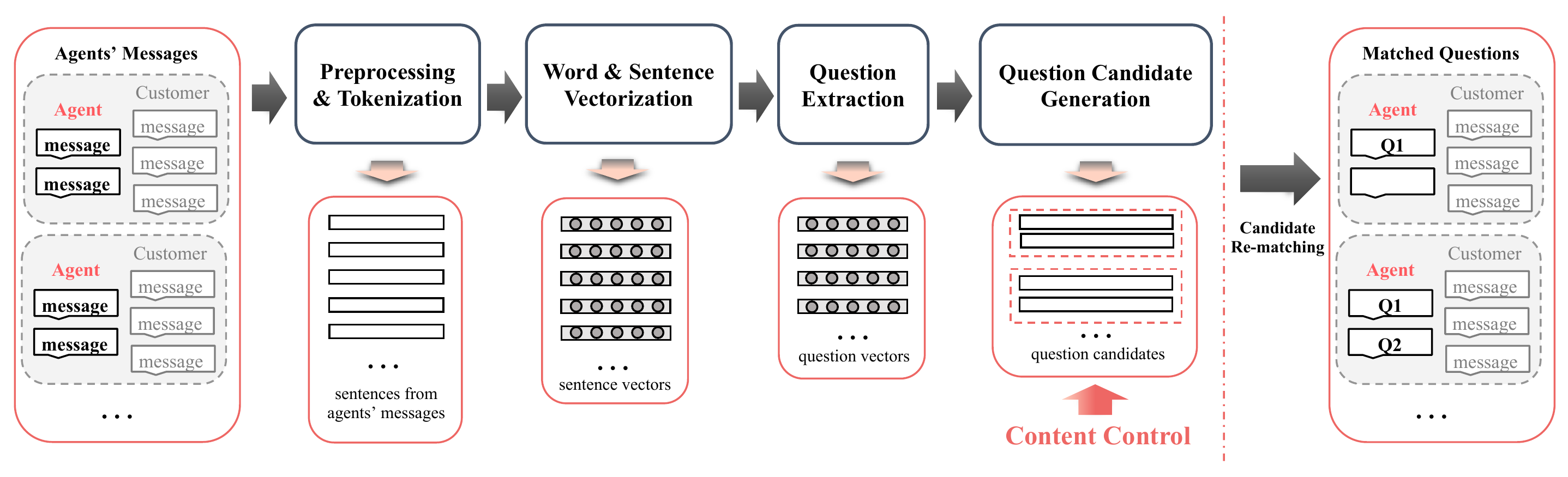}
	\caption{\small Investigative Question Candidate Generation Pipeline.}
\end{figure*}

An important observation from our agent-customer conversation data is that agents' investigation intents can be carried out by asking questions to customers. This motivates us to start inspecting the questions within agents' messages and generating investigative query candidates from there.
We describe the complete question candidate generation pipeline as follows.
\begin{itemize}
	\item \textbf{Preprocessing and Tokenization.} We begin preprocessing by anonymizing personal information in the original conversation messages. This kind of information includes customers' names, agents' names, phone numbers, emails, credit card numbers, URLs, dollar amounts, timestamps and other sensitive information. Note this information will be replaced by the associated placeholders during training but will be automatically filled back at the inference time according to the ticket information.
	After data anonymization, we split agents' messages into sentences and process each sentence into unigram tokens.
    \item \textbf{Vectorization.} The purpose of this step is to transform the original textual information into numeric vector representations so that the downstream machine learning techniques can be applied. 
    We start with all tokenized sentences from agents' messages in our live-chat conversation data and apply \textbf{word2vec} \cite{mikolov2013distributed,mikolov2013efficient} to obtain word representations\footnote{The dimensionality of word embeddings is set to be 300.}. For each word token, we calculate its term frequency -- inverse document frequency (\textbf{Tf-Idf}) in the complete agent's sentence corpus.
    Then we can generate sentence embeddings by aggregating word embeddings based on their associated \textbf{Tf-Idf} weights.
    \item \textbf{Question Extraction.} Now we narrow down the query candidate generation scope on the sentences ending with question marks. 
    Note although the initial candidates are generated from these questions, they will be expended to cover sentences without question marks later by the re-matching process.
    \item \textbf{Investigative Question Candidate Generation.} By analyzing the extracted agents' questions, we observe three different types of questions in our live-chat conversation data.
    \begin{itemize}
        \item \textbf{Courtesy Questions.} These questions usually appear as greetings (e.g.~``How are you doing today?'') or closure signals (e.g.~``Is there anything else I can help you with?'') in agent-customer live-chat conversations.
        \item \textbf{Status-Checking Questions.} In order to ensure that customers are always engaged in our live-chat problem-solving conversations, agents may need to ask status-checking questions occasionally. A typical example of such questions is ``Are you still there with me?''.
        \item \textbf{Investigative Questions.} Apart from courtesy questions and status-checking questions, we find most of the remaining questions is relevant to investigating and resolving customers' issues. We thus refer these questions as investigative questions.
    \end{itemize}
    
    \begin{figure*}
	\includegraphics[width=\linewidth]{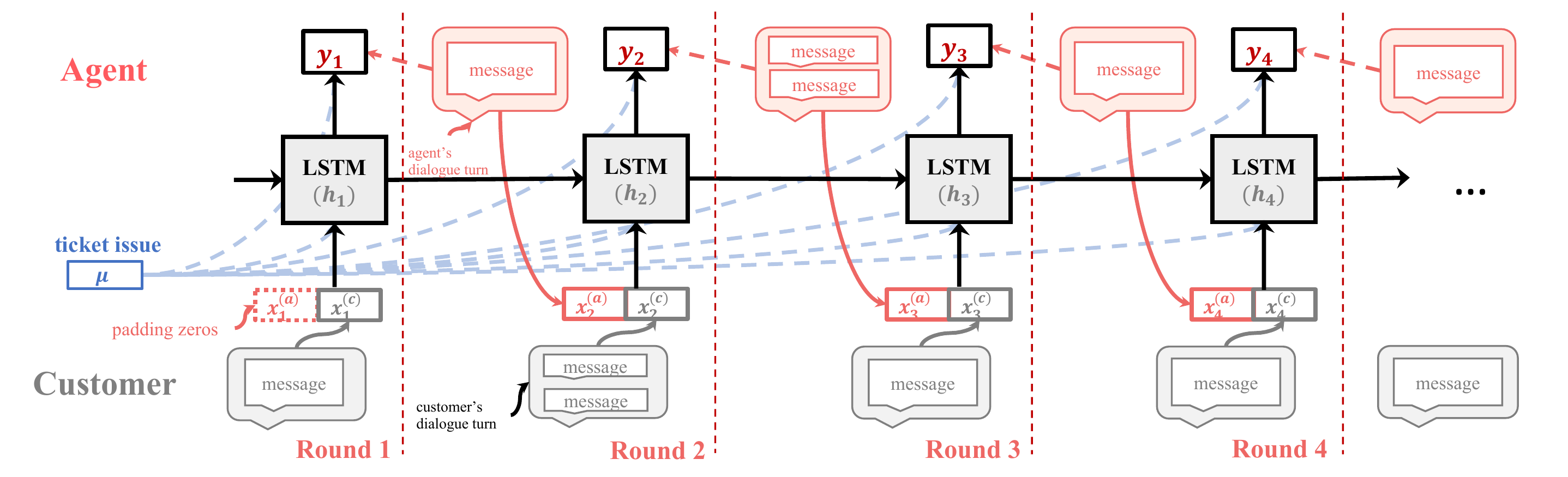}
	\caption{\small Illustration of the response recommendation model architecture for a live-chat conversation with a pre-assigned ticket issue.} \label{fig:model}
\end{figure*}
    
    We notice that variants of agents' courtesy and status-checking questions are relatively limited and can be easily identified by exact matching or simple rules. On the other hand, we still observe more than 60\% of agents' questions are investigative questions, where each with numerous variations and cannot be naively retrieved by exact matching or simple rules. We thus apply a pipeline to cluster these questions and uncover different question variations.
    \begin{itemize}
        \item We first identify several most common courtesy and status-checking questions. Then we remove these questions and their variations, according to the cosine similarities between precomputed sentence embeddings and the courtesy/status-checking question embeddings, from our question candidate pool.
        \item In order to cover investigative question candidates for different ticket issues, for each of the top ticket issues, we extract the associated questions, apply a scalable mini-batch K-means algorithm and assign them into 100 clusters. From each mini-cluster, we extract 3 most frequent variations as candidates. After de-duplicating, we can obtain coarse-grained investigative question candidates.
        \item Then we apply a hierarchical clustering algorithm on the extracted coarse-grained candidates. We refine the top candidates in each cluster and then generate a fine-grained question candidate list.
        \item In order to ensure the high quality of suggested responses, it is important to have our content experts involved in carefully reviewing these response candidates. Reviewing these response candidates is a much lighter and scalable project for the content team compared with writing all potential responses from scratch, which might run into the situation of not applicable to particular contexts, as well as cumbersome to maintain and to keep up to date. In this process, we need to ensure that each candidate cluster represents distinguished semantic intent (e.g.~``Can you provide the reservation code?'' versus ``Do you perhaps have a duplicated account with us?''), different variants can be generated within each cluster (e.g.~``Can you provide the reservation code?'' versus ``Would you mind providing the reservation code?''), and all the response variants are in appropriate tones and styles. After this content control, we are able to finalize 71 distinguished investigative response candidates with around 400 variants in total.
    \end{itemize}

\end{itemize}
Once the investigative question candidates are obtained, we apply a candidate re-matching process to construct a structured training dataset. Specifically, we inspect the complete agents’ sentence corpus, calculate the cosine similarity between each sentence and each question candidate, and assign sentences where the embedding-based cosine similarity is above a given threshold to the closest candidate cluster. Note that all sentences in agents' messages are considered in this re-matching process regardless of question mark, the original investigative question candidates thus can be expended to include their non-question variations (e.g.~from ``Can you provide the reservation code?'' to ``Please let me know the reservation code if you have.'').

\subsection{Investigative Question Recommendation}
After the candidate generation and re-matching process, we are able to approach the investigative response recommendation task.

\begin{table*}
	\centering
	\small
	\begin{tabular}{p{0.1in}p{2.2in}p{2.9in}p{1.1in}}
		\toprule
		 & \textbf{Conversation Reference} & \textbf{Top 3 Recommendations} & \textbf{Evaluation}\\
	   \midrule
		1 & \textbf{Customer:} \textit{The payment link I received by email does not guide me to a payment page. I like to complete this payment. Please help.}  & NA & \textit{NA} \\
		\midrule
		2 & \textbf{Agent:} \textit{Hi \{user name\}. \color{red}{Is this for your reservation with \{host name\}?}}  & \multirow{2}{*}{\makecell[l]{\color{red}1. Could you provide the reservation code or the host name? \\ 2. Which payment method would you like to use?\\3. Have you added the payment method?}} & \textit{Correct} \\
		 & \textbf{Customer:} \textit{Yes.} & & \\
		\midrule
		3 & \textbf{Agent:} \textit{Are you not able to submit the payment through your account as well?} & \multirow{2}{*}{\makecell[l]{1. Have you added the payment method? \\
		    2. Which payment method would you like to use?\\
            3. Did you want to charge your card ending in \{4 digits\}?}} & \textit{Missed} \\
		& \textbf{Customer:} \textit{Not sure how to do that.} & &\\
		\midrule
		4 & \textbf{Agent:} \textit{If you go to your trips tab and click on this reservation, there should be a button on that page to complete the payment.} &  \multirow{2}{*}{\makecell[l]{1. Have you added the payment method? \\
		    2. Which payment method would you like to use? \\
            \underline{3. Are you on the app or website?}}} & \textit{The 3rd recommendation reflects correct context} \\
		& \textbf{Customer:} \textit{Thats empty.} & & \\
		\midrule
	    5 & \textbf{Agent:} \textit{\color{blue}{Are you on the app or website?}} & \multirow{2}{*}{\makecell[l]{\color{blue}1. Are you on the app or website?\\
		    2. Have you added the payment method? \\
            3. Which payment method would you like to use? }} & \textit{Correct}\\
		& \textbf{Customer:} \textit{App.} \\ & & \\
		\bottomrule
	\end{tabular}
	\caption{\small An example agent-customer live-chat dialogue from our test dataset.} \label{table:example}
\end{table*}

\subsubsection{Problem Formulation} Suppose that we have $T$ rounds in the conversation of a customer service ticket $i$, and each conversation round $t$ represents two consecutive 
dialogue turns $\mathcal{T}_{i,t}^{(a)}, \mathcal{T}_{i,t}^{(c)}$ from agent $a$ and customer $c$ respectively. Within each dialogue turn, there could be more than one message sent from the same interlocutor. 
As shown in \cref{fig:model}, an inbound customer service conversation always starts with the customer turn, with a pre-assigned ticket issue. Thus in the first round of a conversation, we have the messages from customer only. 
In the case that the associated ticket issue is missing, we will introduce an additional binary variable to encode this status. Therefore, we could conclude the investigative response recommendation task as follows:
\begin{quote}
    \textbf{Goal:} For a live-chat conversation, given the ticket issue and historical messages from the customer and the agent, we wish to suggest the most likely investigative queries for the agent to ask in the next round of the conversation.
\end{quote}

\subsubsection{Embedding Conversation Turns} Similar to the previous candidate generation process, we apply \textbf{word2vec} on the entire conversation corpus to generate word representations. For each conversation turn, we merge all the messages and aggregate word embeddings based on their associated \textbf{Tf-Idf} weights and obtain turn embeddings. 

\subsubsection{Encoding Structured Outputs} On the other hand, by re-matching the query candidates to the original conversation messages, we are able to construct structured outputs $\bm{y}_{i,t}=(y_{i,t,1}, ..., y_{i,t,m})$ (i.e.,~the true investigative questions that agents asked afterwards) in each conversation round, where $y_{i,t,j}=1$ indicates the $j$-th candidate is hit by the $t+1$-th round of the agent's messages (i.e.,~messages in the next round, see \cref{fig:model}). Notice that agents may ask multiple investigative queries in the same conversation turn. Thus there could be more than one non-zero element in the output vector $\bm{y}_{i,t}$.
At the end, we aim to predict the following conditional probability for the $t$-th conversation round of ticket $i$ \footnote{All the notations are ticket-dependent, for brevity we omit the ticket subscription $i$ in the subsequent contents.}
\begin{equation}
    p_{t,j}\coloneqq P(\underbrace{y_{t,j}=1}_{\mathclap{\text{the $j$-th candidate is hit}}}~|~\overbrace{\bm{x}_{t}^{(a)}, \bm{x}_{t}^{(c)}}^{\mathclap{\text{short-term contexts}}}, \underbrace{\bm{\mu}}_{\mathclap{\text{pre-assigned ticket issue}}}, \overbrace{\bm{x}_{<t}^{(a)}, \bm{x}_{<t}^{(c)}}^{\mathclap{\text{long-term contexts}}}), \label{eq:prediction}
\end{equation}
where $\bm{\mu}$ is a one-hot vector to represent the pre-assigned ticket issue for ticket $i$,  $\bm{x}_{t}^{(a)}$, $\bm{x}_{t}^{(c)}$ are precomputed turn embeddings for the dialogue turns $\mathcal{T}_{t}^{(a)}$ and $\mathcal{T}_{t}^{(c)}$ respectively.

\subsubsection{Model Architecture} 
As shown in \cref{fig:model}, we apply a Recurrent Neural Network (\textbf{RNN}) with Long Short-Term Memory (\textbf{LSTM}) units to model the above probability. We consider a base version of our long-term context model,
where the concatenated dialogue turn embeddings $\bm{x}_{t}=[\bm{x}_{t}^{(a)}; \bm{x}_{t}^{(c)}]$ of each round
are regarded as the input sequence, and the labels $\bm{y}_{t}$ reflected in the next round are regarded as the output sequence. 
The dialog turn embeddings are gradually read in to represent short-term conversation contexts. In addition, the long-term problem contexts are encoded as hidden states $\bm{h}_{t}$ which could be carried and adapted throughout the entire conversation. 
A softmax function $p_{t,j}=\frac{\exp(\bm{w}_j^T\bm{h}_t)}{\sum_{j'} \exp(\bm{w}_{j'}^T\bm{h}_t)}$ can be applied in the output layer to approximate the probability in \cref{eq:prediction}.

To extend the previous base version, the pre-assigned ticket issue $\bm{\mu}$ can be regarded as an additional feature of each ticket $i$, which can be easily concatenated in the input layer with dialogue turn embeddings $\bm{x}_t$ or/and in the output layer with the output hidden states $\bm{h}_t$.

\subsubsection{Training} Our training objective is to maximize the following cross-entropy function
\begin{equation}
    \sum_{i, t}\sum_{j} y_{i,t,j} \log p_{i,t,j}.
\end{equation}
We apply a standard $\ell_2$ regularizer and a stochastic gradient-based method \textbf{ADAM} \cite{kingma2014adam} for optimization. The hyperparameter for the $\ell_2$ regularizer is selected by grid search based on the validation performance.

\subsubsection{Inference} At inference time, we feed the conversation message turns as the input sequence and obtain the output probability distribution over investigative query candidates at each timestamp. We then rank candidates based on their associated probabilities and suggest the top three candidates to agents in the live-chat conversation.

\section{Experiments}

\subsubsection{Data}
We conduct offline experiments on around 20 thousand live-chat ticket conversations and 1 million messages from Airbnb's customer service system. Training, validation and test sets are created using an 80/10/10 random split of these tickets. 

\subsubsection{Baselines}
We consider the following different methods for the investigative query recommendation task.
\begin{itemize}
	\item \textbf{Issue-Wise Frequency.} Query candidates can be simply ranked based on their overall frequencies in the training tickets with the same pre-assigned ticket issue. This straightforward approach does not require additional training efforts and the recommendation outcomes remain static in the entire conversation.
	\item \textbf{Short-Term Context Models.} We remove the \textbf{RNN} architecture in the proposed model and consider messages in the preceding conversation round as short-term contexts only. Specifically, we consider the following two variations.
	\begin{itemize}
	    \item \textbf{Base Version.} We concatenate the agent's and the customer's dialogue turn embeddings and apply a linear model on top of them.
	    \item \textbf{Include ticket issue.} We consider the pre-assigned ticket issue as an additional feature and concatenate it with short-term dialogue turn embeddings. Similarly, a linear model is applied on these features.
	\end{itemize}
	\item \textbf{Long-Term Context Models.} We now consider different variations of the proposed \textbf{RNN} architecture:
	\begin{itemize}
	    \item \textbf{Base Version.} Only dialogue turn embeddings are included in the model.
	    \item \textbf{Include Issue in the Input Layer.} The pre-assigned ticket issue is concatenated with dialogue turn embeddings $\bm{x}_t$ in the input layer.
	    \item \textbf{Include Issue in the Output Layer.} The pre-assigned ticket issue is concatenated with hidden states $\bm{h}_t$ in the output layer.
	    \item \textbf{Include Issue in the Input and Output Layer.} The ticket issue is included in both the input layer with $\bm{x}_t$ and the output layer with $\bm{h}_t$.
	\end{itemize}
\end{itemize}
By comparing short-term context models and long-term context models, we evaluate if capturing long-term problem contexts helps with recommending investigative queries. By comparing different variations of each method, we evaluate if incorporating pre-assigned ticket issues is useful in terms of query recommendation performance.

\subsubsection{Evaluation Metrics}
 We consider Recall@Top3, the coverage of agent's real queries by the suggested top three queries, as the quantitative evaluation metric. Suppose $\hat{\mathcal{Y}_t}$ denote the top three predicted candidates in the round $t$ and $\mathcal{Y}_t$ denote the set of candidates covered in the agent's message. Then our evaluation metric can be defined as $$\mathit{recall}=|\hat{\mathcal{Y}_t}\cap\mathcal{Y}_t|/\min(|\mathcal{Y}_t|, |\hat{\mathcal{Y}_t}|).$$
 We evaluate the recommendation performance on the conversation rounds where at least one query candidate is included in agents' messages. Two average values of such a metric for each method are reported: 1) we compute the average of Recall@Top3 across all valid conversation rounds (\textbf{Recall-r}); and 2) for each ticket conversation, we first calculate the average value across all valid rounds, then compute the average across all the tickets (\textbf{Recall-t}).
 
 \subsubsection{Quantitative Results}
 
 \begin{table}
	\centering
	\begin{tabular}{rcc}
		\toprule
		\textbf{Method} & \textbf{Recall-r} & \textbf{Recall-t} \\
		\midrule
		\textbf{Issue-Wise Frequency } & 0.627 & 0.631 \\
		\midrule
		\textbf{Short-Term Linear} & & \\
		base & 0.646 (0.001) & 0.652 (0.002) \\
		+ issue & 0.648 (0.001) & 0.654 (0.001) \\
		\midrule
		\textbf{Long-Term LSTM} & & \\
		base & \underline{0.689} (0.001) & \underline{0.690} (0.002) \\
		+ issue (in. layer) & 0.688 (0.001) & 0.689 (0.001) \\
		+ issue (out. layer) & 0.688 (0.002) & \underline{0.690} (0.002) \\
		+ issue (in. + out. layers) & 0.686 (0.002) & 0.687 (0.002) \\
		\bottomrule
	\end{tabular}
	\caption{\small Investigative question recommendation results from different methods. Associated standard errors are included in parentheses. Best results are \underline{underlined}.} \label{table:results}
\end{table}

 We run each method 10 times with randomized initializations and report 
 performance of the above models on the investigative question recommendation task. Results are provided in \cref{table:results}. 
 
 From this table we observe that long-term context models significantly outperform short-term contexts models, which indicates that incorporating long-term problem contexts does benefit the performance of recommending investigative responses to agents. In addition, we find incorporating the pre-assigned ticket issue as an additional feature provides a performance boost for the short-term context model. However, such improvements are not significant for the long-term \textbf{LSTM}-based model. A possible reason is that ticket issue information has been implicitly included in the long-term conversation contexts. Therefore, we cannot observe a significant performance gain by explicitly modeling these signals as additional information. This is actually a desired property in real-world applications, as pre-assigned ticket issues could be noisy (customers and agents may choose wrong issues when initializing tickets) or unavailable at beginnings of conversations. Fortunately, such a long-term context model indeed provides us an option to bypass these noisy signals and utilize the conversation messages directly without scarifying recommendation performance.

 \subsubsection{Qualitative Results}
 In \cref{table:example}, the first 5 rounds of one conversation are shown. The customer was having trouble figuring out how to make a payment. From the recommendations for round 3, we can see that the model mistakenly recommended change payment method related investigative questions. As the conversation progressed, the model was able to capture the correct context and start to recommend debugging related investigative questions (e.g., ``Are you on the app or website?''). This demonstrated the effectiveness of the model in capturing both short-term and long-term contexts as the conversation progresses, and correct mistakes with more information.  

\section{Conclusion and Future Work}
In this work, we studied an end-to-end investigative question suggestion system for live-chat customer service agents. We described the pipeline for investigative question candidate generation and proposed a proactive response recommendation model with long-term context memory. 

We leave several online serving challenges as open questions. For example, given that most machine learning infrastructures assume models are stateless, where and how to cache hidden states of the \textbf{RNN} model could be challenging. 
Notice that in our problem setting, multiple messages in the same dialogue turn would be concatenated in the offline training. However, as 
the end-of-a-turn signals are not always accessible in real-time, how to address and resolve this problem during online serving would be worthy of interest as well.

\section{Acknowledgement}
The authors would like to thank Yashar Mehdad and Negin Nejati for helpful discussions and insights, as well as Lisa Qian for her guidance and support.  

The authors would also like to thank members of the Airbnb Agent Platform Design, Engineering, Product, and Content teams for making this work possible:
Tyler Townley, Colleen Purdy, Jen Wardwell, Adrien Cahen, Ted Hadjisavas,  Virginia Vickery, and many others. 

\newpage
\bibliographystyle{aaai}
\bibliography{deepdial}

\end{document}